\documentclass[twocolumn,superscriptaddress,showpacs,amssymb,preprintnumbers,prl,floatfix]{revtex4-1}
\usepackage{epsfig,dcolumn}
\usepackage{hyperref}
\usepackage{multirow}
\usepackage{bm,color}
\usepackage{tikz}
\usepackage{makecell}
\usepackage{graphicx,bm}
\newcommand{\be}{\begin{equation}}
\newcommand{\ee}{\end{equation}}
\newcommand{\ba}{\begin{eqnarray*}}
\newcommand{\ea}{\end{eqnarray*}}

\usepackage{ulem}

\usepackage{soul}
\usepackage{nuc}
\usepackage{tabularx}
\usepackage{booktabs}

\usepackage{array}

\begin{document}
\title{$^{61}$Cr as a Doorway to the $N=40$ Island of Inversion }

\author{L.~Lalanne}
    \email[]{louis.lalanne@iphc.cnrs.fr}
    \affiliation{KU Leuven, Instituut voor Kern- en Stralingsfysica, B-3001 Leuven, Belgium }
    \affiliation{CERN, CH-1211 Geneva 23, Switzerland}
    \affiliation{Université de Strasbourg, CNRS, IPHC UMR 7178, F-67000 Strasbourg, France}

\author{M. Athanasakis-Kaklamanakis}
    \affiliation{KU Leuven, Instituut voor Kern- en Stralingsfysica, B-3001 Leuven, Belgium }
    \affiliation{CERN, CH-1211 Geneva 23, Switzerland}

\author{D.D.~Dao}
  \affiliation{Université de Strasbourg, CNRS, IPHC UMR 7178, F-67000 Strasbourg, France}

\author{Á. Koszorús}
    \affiliation{KU Leuven, Instituut voor Kern- en Stralingsfysica, B-3001 Leuven, Belgium }

 \author{Y. C. Liu}
\affiliation{School of Physics and State Key Laboratory of Nuclear Physics and Technology, Peking University, Beijing 100871, China}

\author{R. Mancheva}
    \affiliation{CERN, CH-1211 Geneva 23, Switzerland}
    \affiliation{KU Leuven, Instituut voor Kern- en Stralingsfysica, B-3001 Leuven, Belgium }

\author{F.~Nowacki}
  \affiliation{Université de Strasbourg, CNRS, IPHC UMR 7178, F-67000 Strasbourg, France}

  \author{J. Reilly}
\affiliation{Department of Physics and Astronomy, The University of Manchester, Manchester M13 9PL, United Kingdom}

\author{C. Bernerd}
   \affiliation{CERN, CH-1211 Geneva 23, Switzerland}

\author{K. Chrysalidis}
   \affiliation{CERN, CH-1211 Geneva 23, Switzerland}

\author{ T. E. Cocolios}
    \affiliation{KU Leuven, Instituut voor Kern- en Stralingsfysica, B-3001 Leuven, Belgium }

\author{R. P. de Groote}
    \affiliation{KU Leuven, Instituut voor Kern- en Stralingsfysica, B-3001 Leuven, Belgium }

\author{K. T. Flanagan}
\affiliation{Department of Physics and Astronomy, The University of Manchester, Manchester M13 9PL, United Kingdom}

\author{R. F. Garcia Ruiz}
\affiliation{Massachusetts Institute of Technology, Cambridge, MA 02139, USA}

\author{D. Hanstorp}
\affiliation{Department of Physics, University of Gothenburg, SE-412 96 Gothenburg, Sweden}

\author{R. Heinke}
    \affiliation{KU Leuven, Instituut voor Kern- en Stralingsfysica, B-3001 Leuven, Belgium }

\author{M. Heines}
    \affiliation{KU Leuven, Instituut voor Kern- en Stralingsfysica, B-3001 Leuven, Belgium }

\author{P. Lassegues}
    \affiliation{KU Leuven, Instituut voor Kern- en Stralingsfysica, B-3001 Leuven, Belgium }

\author{K. Mack}
\affiliation{Department of Physics and Astronomy, The University of Manchester, Manchester M13 9PL, United Kingdom}

\author{B. A. Marsh}
   \affiliation{CERN, CH-1211 Geneva 23, Switzerland}

\author{A. McGlone}
\affiliation{Department of Physics and Astronomy, The University of Manchester, Manchester M13 9PL, United Kingdom}

\author{K. M. Lynch}
\affiliation{Department of Physics and Astronomy, The University of Manchester, Manchester M13 9PL, United Kingdom}

\author{G. Neyens}
    \affiliation{KU Leuven, Instituut voor Kern- en Stralingsfysica, B-3001 Leuven, Belgium }

\author{B. van den Borne}
    \affiliation{KU Leuven, Instituut voor Kern- en Stralingsfysica, B-3001 Leuven, Belgium }

\author{R. Van Duyse}
    \affiliation{KU Leuven, Instituut voor Kern- en Stralingsfysica, B-3001 Leuven, Belgium }

\author{X. F. Yang}
\affiliation{School of Physics and State Key Laboratory of Nuclear Physics and Technology, Peking University, Beijing 100871, China}

\author{J. Wessolek}
\affiliation{Department of Physics and Astronomy, The University of Manchester, Manchester M13 9PL, United Kingdom}
    \affiliation{CERN, CH-1211 Geneva 23, Switzerland}

\begin{abstract}
This paper reports on the measurement of the ground-state spin and nuclear magnetic dipole moment of $^{61}$Cr. The radioactive ion beam was produced at the CERN-ISOLDE facility and was probed using high-resolution resonance ionization laser spectroscopy with the CRIS apparatus. The present ground-state spin measurement  $I = \frac{1}{2}$, differing from the previously adopted $I =(\frac{5}{2})$, has significant consequences on the interpretation of existing beta decay data and nuclear structure in the region. The structure and shape of $^{61}$Cr is interpreted with state-of-the-art Large-Scale Shell-Model  and Discrete Non-Orthogonal Shell-Model calculations. From the measured magnetic dipole moment $\mu(^{61}$Cr$)=+0.541(6)~\mu_N$ and the theoretical findings, its configuration is understood to be driven by 2 particle - 2 hole neutron excitations with an unpaired $1p_{1/2}$ neutron. This establishes the western border of the $N=40$ Island Of Inversion, characterized by 4 particle \text{-} 4 hole neutron components. We discuss the shape evolution along the Cr isotopic chain as a 2$^{nd}$ order quantum phase transition at the entrance of the $N=40$ Island of Inversion.

 \end{abstract}

%\pacs{PACS number(s): 21.60.Cs, 23.40.-s, 21.10.-k, 27.40.+z}
%\pacs{21.10.--k, 27.40.+z, 21.60.Cs, 23.40.--s}
\keywords{Neutron rich nuclei, Shell Model, Shell Evolution, Quantum Phase Transition, Laser Spectroscopy, Nuclear moments.}

\date{\today}
\maketitle

\noindent {\sl Introduction.} 
Understanding how nuclear structure evolves at large proton-to-neutron asymmetries is one of the main quests of contemporary nuclear physics.
The disappearance of a shell closure far from stability is generally accompanied by an increase in configuration mixing and collectivity, thus leading to the formation of Islands of Inversion  \cite{Sor08,Ots20,Now21}. In an Island of Inversion, strong quadrupole correlations energetically favor deformed intruder configurations ($i.e.$ multiparticle, multihole $\mathit{Np}\text{-}\mathit{Nh}$ cross-shell excitations) which hence become the ground-state \cite{Hey11,Cau14,Len10}.
The nuclear shell model has guided our understanding of the emergence of Islands of Inversion in terms of intruder states in the past decades. However, capturing the full extent of shape evolution associated with the formation of an Island of Inversion requires a different theoretical framework. The recently developed Discrete Non-Orthogonal Shell-Model (DNO-SM)~\cite{Dao22} combined with state-of-the-art Large Scale Shell Model (LSSM) calculations provide a powerful approach to link the microscopic structural changes to the macroscopic shape evolution, occurring during the formation of the Islands of Inversion.

Experimentally, four Islands of Inversion are known to this date, associated with the neutron (sub-)shell closures at $N=$ 8, 20, 28, and 40 \cite{Now21}. 
In the $N=40$ Island of Inversion, collectivity rapidly develops below $^{68}$Ni, as a result of a narrowing of the $N=40$ subshell gap while emptying the proton $f_{7/2}$ orbital \cite{Len10,Lju10}. The chromium isotopes ($Z=24$) exhibit the strongest level of deformation in the region. Along this isotopic chain, mass measurements as well as the $2^+$ excitation energies and transition probabilities reveal a steep increase in quadrupole correlations between $N=32$ and $N=38$ and strongly deformed ground-state for $^{60,62,64}$Cr \cite{Sor03, Bau12,Mou18}. 
While the deformation is suggested to be nearly constant from $^{62}$Cr onward, with a maximum achieved at $N=40$ \cite{Gade21, San15}, recent mass measurements point towards a more rapid change in correlation effects in $^{63}$Cr, towards $N=40$ \cite{Sil22}. $^{62,64}$Cr are understood to be members of the $N=40$ Island of Inversion and to share similar configurations dominated by $\mathrm{4}\mathit{p}\text{-}\mathrm{4}\mathit{h}$ neutron excitations across $N=40$ \cite{Cra13,Gade21}, with several competing shapes identified in the low-energy structure of $^{62}$Cr \cite{Gad24}.
Despite the progress made in understanding this structural evolution, questions remain about the exact nature and location of the transition into the $\mathrm{4}\mathit{p}\text{-}\mathrm{4}\mathit{h}$ regime of the $N=40$ Island of Inversion. The structure and shape of $^{61}$Cr ($N=37$) remains to be interpreted  to  better understand how the $N=40$ Island of Inversion forms.

Despite the interest, experimental information about $^{61}$Cr is scarce and its structure is poorly understood. While its mass \cite{Mou18} and lifetime \cite{Ame98,Sor99} are known, its spin-parity $I^\pi$ has been only tentatively assigned to be $(\frac{5}{2}^-)$ by beta decay studies \cite{Gau05,Cra09,Suc14}. The high density of states at low energy and the nearly-degenerate neutron $f_{5/2}$ and $p_{1/2}$ orbitals in the region, make spin assignment and structure interpretation difficult both experimentally and theoretically. Furthermore, building a reliable level scheme, which is crucial to understand the excited structure of the nucleus, is impossible without measuring the ground-state spin $I$ and magnetic dipole moment $\mu$.

This letter reports on the first high-resolution laser spectroscopy of $^{61}$Cr. Its ground-state $I$ and $\mu$ are determined in a nuclear model-independent way, establishing its wave function properties. Its structure, shape, and $\mathit{Np}\text{-}\mathit{Nh}$ content are interpreted with the combination of state-of-the-art LSSM and DNO-SM calculations, in the context of the transition into the $N=40$ Island of Inversion. An abrupt change in the microscopic structure of the ground state of chromium isotopes is observed, accompanied by a shape transition. This change is identified as a quantum phase transition, occurring at the entrance of the $N=40$ Island Of Inversion.

%%%%%%%%%%%%%%%%%%%%%%%%%%%%%%%%%%%%%%%%%%%%%%%%%%%%%%%%%%%%%%%%

\noindent {\sl Experimental techniques.} 
The chromium ion beams were produced at the CERN-ISOLDE facility \cite{ISOLDE}. Fission fragments were produced by impinging a pulsed, 1.4-GeV proton beam on a thick uranium-carbide target. After diffusing out of the target, the Cr atoms were ionized using the Resonant Ionization Laser-Ion Source (RILIS) \cite{RILIS}. Ions were then accelerated to 30 keV and mass-selected using a high-resolution separator. The beam was cooled and bunched in a gas-filled linear Paul trap \cite{ISCOOL}, before it was sent to the Collinear Resonance Ionization Spectroscopy (CRIS) beam line \cite{CRIS,VoltageScaning}. The $^{61}$Cr beam, at 3000 ions/s, had about 2$\times 10^5$ ions/s of isobaric contamination, mainly from stable molecular compounds such as $^{42}$Ca$^{19}$F or $^{45}$Sc$^{16}$O. 

\begin{figure}[!h]
\begin{center}
\centering
\includegraphics[width=0.99\columnwidth]{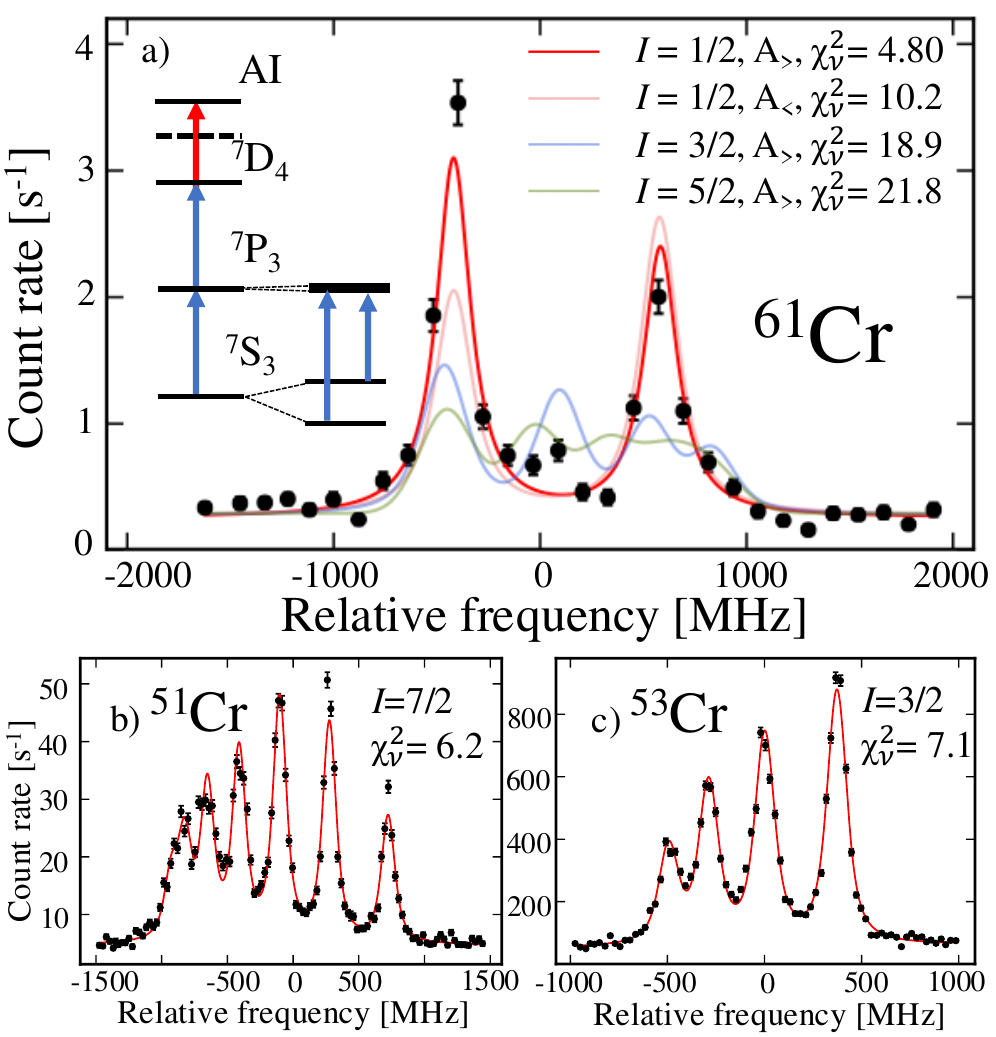}
 \end{center}
\caption{Hyperfine spectrum of a) $^{61}$Cr, b) $^{51}$Cr and c) $^{53}$Cr: count rate in the ion detector as a function of the frequency of the first excitation step. Frequencies are given relative to the centroid of the hyperfine structure. Black markers are experimental data while colored lines represent fits using different nuclear spin $I$ hypothesis.  The inset in a) shows a schematic of the three-step laser ionization scheme. A$_{>}$ (A$_{<}$) indicates if the $A_{S_3}$ parameter is positive (negative) in the fit.  $\chi^2_\nu$ is the reduced chi-square per degree of freedom. }
\label{fig:hfs}
\end{figure}

In the CRIS setup, the ion beam was firstly neutralized using a charge-exchange cell (CEC) filled with Na vapor at 210(5)$^\circ$C.  
The remaining ions  were dumped using a set of electrostatic deflectors while the atom bunches were sent to the interaction region (maintained at $2\times10^{-9}$~mbar to minimize non-resonant collisional ionization), where they were collinearly overlapped in space and time with three laser pulses to resonantly excite and ionize the Cr atoms. The laser-ions were then deflected towards a MagneToF single-ion detector.
The frequency of the first laser step (inset of 
Fig.~\ref{fig:hfs}a)) was scanned in the ion's rest-frame by tuning the velocity of the ions using a voltage scanning device \cite{VoltageScaning} placed before the CEC. 

The inset in Fig.~\ref{fig:hfs}a) shows a schematic of the laser ionization scheme used in this work. The first atomic transition at 427.48~nm $3d^5(^6S)4s$ $^7S_3 \rightarrow 3d^5(^6S)4p$ ${^7P_3}$, was probed by laser light from an injection-locked
 titanium:sapphire (Ti:Sa) cavity (20~MHz linewidth), seeded by a narrowband, continuous-wave Ti:Sa laser (Sirah Matisse). Atoms in the excited ${^7P_3}$ state were further excited to the $3d^44s5s$ ${^7D_4}$ atomic state by an intra-cavity doubled grating Ti:Sa  (6(1)~GHz linewidth) exciting the 427.29-nm transition. Finaly, the Cr atoms were ionized by populating an auto-ionizing state (located 4948 cm$^{-1}$ above the ionization potential), using a 785-nm transition driven by a single etalon Ti:Sa cavity (12(2)~GHz linewidth). The three pulsed Ti:Sa cavities were pumped at 1~kHz by a 532-nm Nd:YLF laser. Laser frequencies were measured with a HighFinesse WS/U-2 wavelength-meter, with a $10^{-8}$ relative accuracy. 
 
 With the present Ti:Sa-only ionization scheme, developed in this work to be used both for production at RILIS and for high-resolution spectroscopy at CRIS, the overall efficiency of the experiment was improved by an order of magnitude, compared to previous schemes.
 Furthermore, thanks to the selectivity of the present scheme, a maximum total efficiency of 1 per 300 was reached for even-even chromium isotopes, while suppressing by six orders of magnitude the molecular contaminants.
Without this new scheme, background arising from this contamination would have been too high for the present measurement to be performed.

\noindent {\sl Results.} 
Examples of the measured hyperfine structures (HFS) of $^{61}$Cr and of the two benchmark nuclei $^{51}$Cr and $^{53}$Cr are shown in Fig.~\ref{fig:hfs}. The background is dominated by non-resonant collisional ionization of chromium isotopes (for mass 51 and 53) or stable molecular contaminants (for mass 61). The error bar of each data point corresponds to the statistical uncertainty. 
The upper excited state ${^7P_3}$ has hyperfine $A_{P_3}$ and $B_{P_3}$ parameters well below the present experimental resolution ($FWHM\approx300$~MHz for $^{61}$Cr and 70~MHz for $^{51,53}$Cr) \cite{Bec78, Buc66, Rei95}. 
Thus it is expected, as long as the nuclear ground-state spin $I$ is smaller than the atomic ground-state spin $J =3$ \cite{Sug77}, to observe a one-to-one correspondence between $I$ and the observed number of peaks in the HFS, corresponding to the number of hyperfine states $F$, where $\vert I-J\vert<F<I+J$ \cite{Yang23}. 
One can observe in Fig.~\ref{fig:hfs}b) and c) the HFS of $^{51}$Cr ($I=\frac{7}{2}$)  and $^{53}$Cr ($I=\frac{3}{2}$), with seven (leftmost peak being a doublet) and four peaks, respectively.   
The HFS of $^{61}$Cr Fig.~\ref{fig:hfs}a) displays a clear double peak structure which can only be observed with  $I=\frac{1}{2}$. The presence of a possible $\frac{9}{2}^+$ isomeric state in $^{61}$Cr, similar to the isotone nucleus $^{63}$Fe \cite{Lun07,Blo08}, has been investigated but is not observed in our measurements.

The analysis of the HFS was performed using the SATLAS2 Python package \cite{SATLAS2}. The fit function is composed of multiple Voigt functions plus a linear background, with the peak positions determined by the hyperfine Hamiltonian \cite{Yang23}. The fits in  Fig.~\ref{fig:hfs} used fixed relative peak amplitudes based on Racah intensities \cite{Yang23} (except for $^{51}$Cr, Fig.~\ref{fig:hfs}b)).  The $A_{P_3}$ and $B_{P_3}$ hyperfine parameters of the ${^7P_3}$ state  were fixed to zero, as well as the $B_{S_3}$ parameter of the ${^7S_3}$ state (all measured to be compatible with zero within 2~MHz \cite{Rei95, Jar07, Chi63,Bec78, Buc66}). All the other fitting parameters (centroid and amplitude of the HFS, $A_{S_3}$, Gaussian and Lorentzian widths and background parameters) were left free. The red, blue, and green colored lines in Fig.~\ref{fig:hfs}a) shows the best fit to the data considering $I = \frac{1}{2}, \frac{3}{2}$ and $\frac{5}{2}$, respectively. Data can only be reproduced with the $\frac{1}{2}$ spin hypothesis.

The experimental method used for extracting $\mu$ was benchmarked by comparing our measurements of $^{51,53}$Cr, to literature values. The measured hyperfine $A_{S_3}$ parameter for $^{53}$Cr $A_{S_3}(^{53}$Cr$)=-82.64(16)$~MHz, agrees with the literature value $A_{S_3,lit}(^{53}$Cr$)=-82.5985(15)$~MHz \cite{Chi63}. 
The differential hyperfine anomaly \cite{Per23} is measured to be 0.2(15)$\%$ using nuclear magnetic resonance measurements \cite{Stone} and is therefore neglected in the analysis. 
Thus, using the precise value $\mu_{lit}(^{53}$Cr$)=-0.47454(3)~\mu_N$ \cite{Ald53}, the $\mu$ of a Cr isotope of mass $A$ can be determined from $\mu(^A\textnormal{Cr})=\mu_{lit}(^{53}\textnormal{Cr})\frac{A(^{A}\textnormal{Cr})I(^{A}\textnormal{Cr})}{A_{lit}(^{53}\textnormal{Cr})I(^{53}\textnormal{Cr})}$. The measured 
$\mu(^{51}$Cr$)=-0.933(13)~\mu_N$ closely matches the literature value $\mu_{lit}(^{51}$Cr$)=-0.934(5)~\mu_N$ \cite{Stone}, further confirming the accuracy of the method.  Note that, even using a spectroscopic transition with larger $B$ parameters, no electric quadrupole moment can be extracted for $I=0$ or $\frac{1}{2}$.

The  hyperfine $A_{S_3}$ parameter of $^{61}$Cr has been measured to be $A_{S_3}(^{61}$Cr$)=+282.4(28)$~MHz leading to $\mu(^{61}$Cr$)=+0.541(6)~\mu_N$. The value is determined by the relative position of the two peaks in Fig.~\ref{fig:hfs}a) while the sign is constrained by their relative amplitudes. A fit using a negative $A_{S_3}$ (shown in light red in Fig.~\ref{fig:hfs}a)) leads to a reduced chi-square $\chi^2_\nu=10.2$, twice as large as the one obtained with positive $A_{S_3}$.  The value and uncertainty correspond to the $\chi^2$-normalized weighed average and standard deviation of four independent measurements performed over three days. The total error includes all experimental systematic errors, dominated by effects such as the experimental resolution, instabilities of the wavemeter or error on the measurement of the acceleration voltage. The error of a single measurement is taken as the fit error (taking into account statistical errors, experimental resolution and line-shape distortion). The four independent measurements are all compatible within 1 $\sigma$, with a typical uncertainty of 5~MHz.

 The measured magnetic dipole moment $\mu(^{61}$Cr$)=+0.541(6)~\mu_N$ lies close to the single-particle value $+0.638~\mu_N$ of a $\nu p_{1/2}$ state, supporting a negative parity state. Assuming that $I$ and $\mu$ are dominated by the odd neutron, the positive sign of the moment determines a $l-1/2$ coupling of the occupied orbital and excludes the possibility of a $\nu g_{9/2}$, $\nu d_{5/2}$ or $\nu s_{1/2}$  state and therefore of a positive parity. 
Furthermore, there are no $I^\pi=\frac{1}{2}^+$ level observed below 500~keV in the region \cite{NNDC}.
These arguments support a  negative parity and therefore $I^\pi=\frac{1}{2}^-$ for $^{61}$Cr.

\begin{figure}[!ht]
\begin{center}
\centering
\includegraphics[width=0.99\columnwidth]{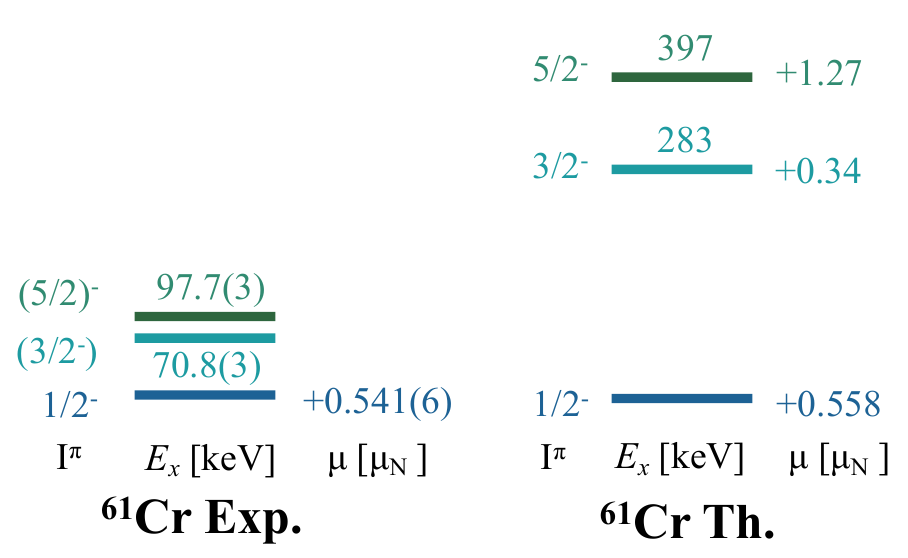}
 \end{center}
\caption{Experimental and theoretical partial level scheme of $^{61}$Cr with spin-parities $I^\pi$, excitation energies $E_x$ and magnetic dipole moments $\mu$. Experimental excitation energies are from Ref. \cite{Suc14}.  }
\label{fig:ExScheme}
\end{figure}

\noindent {\sl Discussion.}    
The ground-state spin-parity of $^{61}$Cr had previously been tentatively assigned $(\frac{5}{2}^-)$,  based on its direct feeding from the beta decay of $^{61}$V \cite{Gau05} (whose spin is still not established).
This assignment was supported by the beta decay study of $^{61}$Cr ~\cite{Cra09}, based on the large feeding to the $\frac{5}{2}^-$ ground-state of $^{61}$Mn (now firmly established \cite{Bad15}) and to its first  $(\frac{7}{2}^-)$ excited state \cite{Val08}. 
The measurement reported here provides the first direct spin measurement of the $^{61}$Cr ground-state, establishing $I^\pi=\frac{1}{2}^-$ and revising the previously adopted $I^\pi=(\frac{5}{2}^-$).

The beta decay of the $\frac{1}{2}^-$ $^{61}$Cr ground-state to the $\frac{5}{2}^-$  $^{61}$Mn ground-state corresponds to a second forbidden transition, contradicting its measured  log~$ft=5.1(2)$ \cite{Cra09}. This discrepancy may stem from the overestimation of ground-state direct beta feeding due to the missing beta strength for excitation energies $E_x>2.5$~MeV, known as the Pandemonium effect \cite{Har77}. 
The situation is similar for the $^{61}$Mn $(\frac{7}{2}^-)$ first excited state which should be weakly populated by direct beta feeding. It would be valuable to investigate these inconsistencies with additional beta decay data, including total absorption spectroscopy. The reported apparent direct beta feeding from $^{61}$V to the ground-state of $^{61}$Cr \cite{Gau05} suggests $I^\pi=(\frac{1}{2}^-)$ or $(\frac{3}{2}^-)$ for the ground-state of $^{61}$V. However, the missing beta strength and the lack of spectroscopy data on $^{61}$V prevent any conclusion on its possible ground-state spin and configuration. Additional spectroscopy data on this nucleus are also required.  

The beta decay study of $^{61}$V identified two $\gamma$ transitions feeding the $^{61}$Cr ground-state from low-lying excited states at 70.8(3) keV and 97.7(3) keV \cite{Suc14}. Based on Weisskopf estimates and the non-observation of a lifetime for these states, it is suggested that the 71~keV transition is limited to a dipole character, while the 98~keV transition would be a dipole or mixed $M1-E2$ transition \cite{Suc14}. Dipole transitions ($M1$ or $E1$) from the $I^\pi=\frac{1}{2}^-$ state would populate $\frac{1}{2}^\pm$ or $\frac{3}{2}^\pm$ states while an E2 transition would populate $\frac{3}{2}^-$ or  $\frac{5}{2}^-$ states. No low-lying $\frac{1}{2}^+$ and $\frac{3}{2}^+$ are observed for even-$Z$ odd-$N$ nuclei in the region, making this hypothesis unlikely. 
Furthermore, a low-lying triplet $\frac{1}{2}^-$, $\frac{3}{2}^-$, $\frac{5}{2}^-$ is observed in several even-$Z$, $N=37$ isotones and odd-$N$, $Z=24$ isotopes, around $^{61}$Cr. Based on these arguments, on the present ground-state $I^\pi=\frac{1}{2}^-$  determination and on the reported multipolarities from Ref.~\cite{Suc14}, the tentative spin-parity assignments of $(\frac{3}{2}^-)$ and $(\frac{5}{2})^-$ are proposed for the 70.8(3) keV and 97.7(3)-keV states, respectively (see Fig. \ref{fig:ExScheme}).

\begin{figure}[!ht]
\begin{center}
\centering
\includegraphics[width=0.99\columnwidth]{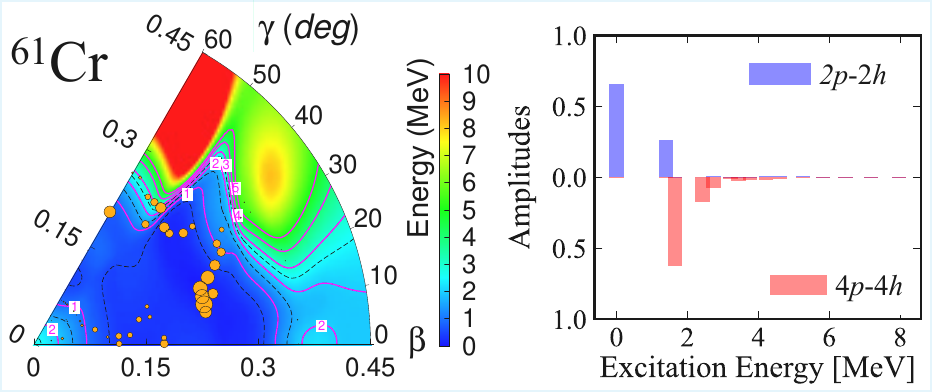}
 \end{center}
\caption{Left: ground-state wavefunction of $^{61}$Cr  expanded in the $(\beta,\gamma)$ plane.  Right: Theoretical amplitudes of the deformed neutron $\mathrm{2}\mathit{p}\text{-}\mathrm{2}\mathit{h}$ (blue) and $\mathrm{4}\mathit{p}\text{-}\mathrm{4}\mathit{h}$ (red) components into all the  $\frac{1}{2}^-$ states.}
\label{fig:61CrTh}
\end{figure}

%%%%%%%%%%%%%%%%%%%%%%%%%%%%%%%%%%%%%%%%%%%%%%%%%%%%%%%

\noindent {\sl Interpretation.} We performed theoretical calculations within LSSM diagonalizations
and further interpreted our results with the recently-developed DNO-SM approach~\cite{Dao22}. Within these frameworks, 
 the valence space  is composed of the $pf$-shell for protons and  
the $1p_{3/2}$, $0f_{5/2}$, $1p_{1/2}$, $0g_{9/2}$, and $1d_{5/2}$ orbitals for neutrons, on top of an inert $^{48}$Ca core.
The interaction is the LNPS Hamiltonian \cite{Len10} with minor adjustments \cite{Now18}, which has shown to be extremely successful for the description of collectivity and the Island of Inversion at $N=40$ 
~\cite{Lju10,Bau12,San15,Gad14,Cor20}.
We recall that using the same valence space, the DNO-SM  allows diagonalization of the same LNPS effective interaction in a relevant deformed Hartree-Fock (HF) states basis from the potential energy surface represented in a $(\beta,\gamma)$ plane. The diagonalization was then performed after rotational symmetry restoration using angular-momentum projection technique. 

The theoretical spectroscopic results are compared to experiment in Fig.~\ref{fig:ExScheme}.
The calculations reproduce the  $I^\pi=\frac{1}{2}^-$ ground-state and support the $\frac{1}{2}^-$ - $\frac{3}{2}^-$ - $\frac{5}{2}^-$ states sequence proposed in this work.  The calculations also produce close-lying $\frac{3}{2}^-$ - $\frac{5}{2}^-$ excited states but slightly overestimate their excitation energies. With two excited states lying within less than 100~keV, $^{61}$Cr displays the highest level density of the isotopic chain at low excitation energies. 

For the calculation of $\mu$, we use bare  orbital $g^{\pi,\nu}_l$ and spin $g^{\pi,\nu}_s$ factors. These bare $g$-factors values (shown in Fig. \ref{fig:ExScheme}) reproduce the $\mu$ measurement within 2$\sigma$ and confirms its $I^\pi=\frac{1}{2}^-$ nature.
It should be noted that in the shell-model framework, effective operators, and in particular spin operators renormalized (quenched) to the model space, should be used to account for external contributions, similar to two-body currents used in \textit{ab-initio} calculations. 
A recent survey on the impact of two-body currents on $\mu$ shows that its contribution is  nucleus-to-nucleus dependent  with sizeable variations, pointing to a probable minor contribution in the present case \cite{Miy24}.
A more general study of these effective operators, beyond the scope of the present paper, should be undertaken in the future.

\begin{figure}[!ht]
\begin{center}
\centering
\includegraphics[width=0.99\columnwidth]{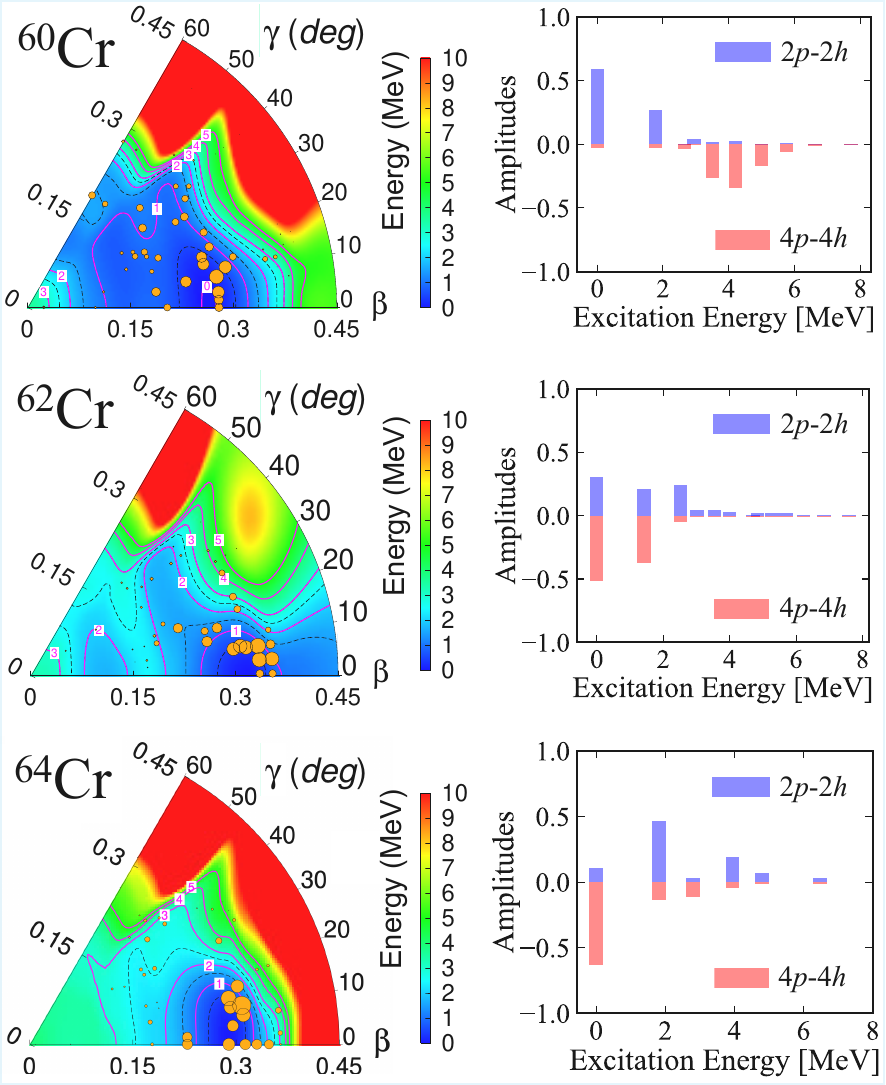}
 \end{center}
\caption{Same as Fig~\ref{fig:61CrTh} for ground-state wavefunctions of $^{60,62,64}$Cr.}
\label{fig:Th}
\end{figure}

Additional insight into the nature of the low-lying states is provided by inspection of the wave function contents. The theoretical occupations support the $I^\pi=\frac{1}{2}^-$ and $\mu$ being carried by an unpaired $\nu p_{1/2}$ neutron. The remaining orbital occupancies reveal an intruder configuration with a strong population of $\nu g_{9/2}$ and $\nu d_{5/2}$ intruder orbitals of 1.65 neutrons, suggesting the deformed character of the state. Transfer reaction data to experimentally confirm these findings would be of high interest.  
In the left panel of Fig.~\ref{fig:61CrTh}, we show the $\frac{1}{2}^-$ wavefunction expanded in the $(\beta,\gamma)$ plane.
It shows strongly-deformed triaxial components  centered around $(\beta,\gamma)$ $\sim$ $(0.23, 20^{\circ})$ of mainly $\mathrm{2}\mathit{p}\text{-}\mathrm{2}\mathit{h}$ nature.
Interestingly, when extracting the overlaps between the deformed neutron $\mathrm{2}\mathit{p}\text{-}\mathrm{2}\mathit{h}$ and $\mathrm{4}\mathit{p}\text{-}\mathrm{4}\mathit{h}$ fixed configurations and the physical states (right panel of the same figure), the $\mathrm{2}\mathit{p}\text{-}\mathrm{2}\mathit{h}$ probability amplitude is dominant in $^{61}$Cr, while the  $\mathrm{4}\mathit{p}\text{-}\mathrm{4}\mathit{h}$ deformed configuration is not present at the location of the ground-state, but mostly concentrated in a single excited state around 2 MeV. 

With one additional neutron, $^{62}$Cr has recently been shown to exhibit a complex low-lying structure with the occurrence of three rotational bands~\cite{Gad24} and its ground-state dominated by  $\mathrm{4}\mathit{p}\text{-}\mathrm{4}\mathit{h}$ configurations, which defines the $N=40$ Island of Inversion. In view of these characteristics, it is therefore interesting to assess
the evolution of the deformation entering the $N=40$ Island of Inversion. For that, we perform the same theoretical treatment for $^{60,62,64}$Cr isotopes. 
One observes, in Fig.~\ref{fig:Th} (left panel), a clear evolution  from a mildly collective wave function  in $^{60}$Cr, to stronger collective and axial regimes in $^{62}$Cr and $^{64}$Cr with an intermediate triaxial regime developing in $^{61}$Cr. Such transitions, where a sudden transformation in the structure of the ground state is observed, varying $N$ or $Z$, are often qualified as quantum phase transitions \cite{PhaseTrans, ZrOtsuka}.

The underlying mechanism associated with these transitions is connected to the $\mathit{Np}\text{-}\mathit{Nh}$ content of the involved wave functions: the $\mathrm{2}\mathit{p}\text{-}\mathrm{2}\mathit{h}$ probability amplitude (right panel) is the strongest in the $^{60}$Cr ground-state but becomes progressively suppressed due to the $\mathrm{4}\mathit{p}\text{-}\mathrm{4}\mathit{h}$ emergence which dominates in the ground-state of $^{62,64}$Cr. 
In \cite{San15}, $^{64}$Cr was discussed to be the most collective chromium isotope in the Island of Inversion at $N=40$ with strongly dominating $\mathrm{4}\mathit{p}\text{-}\mathrm{4}\mathit{h}$ components. From the present scenario, we can  infer that  $^{61}$Cr is the first chromium isotope where $\mathrm{4}\mathit{p}\text{-}\mathrm{4}\mathit{h}$ components appear in the low-lying excited spectrum as well as the last chromium isotope to have a ground-state dominated by $\mathrm{2}\mathit{p}\text{-}\mathrm{2}\mathit{h}$ configurations. This feature makes $^{61}$Cr a key transitional nucleus, placed at the western entrance to the $N=40$ Island of Inversion.
 This structural evolution also reflects the nature of the QPT. First-order QPTs are characterized by an abrupt change, such as the sharp crossing between the 0$_1^+$ and 0$_2^+$ states seen in $^{100}$Zr \cite{, ZrOtsuka}, while second-order QPTs are characterized by a continuous  evolution with a higher level of configuration mixing between a larger number of states \cite{PhaseTrans}. Along the Cr isotopic chain, the transition is not driven by a single crossing but by a gradual mixing of multiple states, as shown in Fig.~\ref{fig:Th} (right panel), characterizing a second order quantum phase transition.

In conclusion, the ground-state spin and nuclear magnetic dipole moment of $^{61}$Cr were measured for the first time using the CRIS experiment at the CERN-ISOLDE facility. Our measurement establishes $I^\pi=\frac{1}{2}^-$ as the spin-parity of $^{61}$Cr, instead of $I^\pi=(\frac{5}{2}^-$) as previously adopted in the literature and is used to re-evaluate the level scheme of  $^{61}$Cr. This measurement also highlights the need of additional decay spectroscopy and transfer reaction data in the region, especially for $^{61}$Cr and $^{61}$V. State-of-the-art LSSM calculations reproduce the experimental findings. The $^{61}$Cr ground-state is interpreted, following the DNO-SM wave-function analysis, to have a triaxial shape and to be dominated by a $\mathrm{2}\mathit{p}\text{-}\mathrm{2}\mathit{h}$ neutron intruder configuration coupled to an unpaired $\nu 1p_{1/2}$ neutron. Within the isotopic chain, $^{61}$Cr is understood to make the transition between the $\mathrm{2}\mathit{p}\text{-}\mathrm{2}\mathit{h}$ and the $\mathrm{4}\mathit{p}\text{-}\mathrm{4}\mathit{h}$ regime. In addition, a fast shape evolution is associated with the microscopic changes of the wave function, interpreted as a second order quantum phase transition at the entrance of the $N=40$ Island of Inversion.

\acknowledgments {We thank the ISOLDE technical teams for their support. L.L. thanks G. Tocabens and H.L. Crawford for fruitful discussions. This project has received funding from the European Union's Horizon Europe research and innovation programme under grant agreement no. 101057511 (EURO-LABS); as well as from FWO International Research Infrastructures project I001323N, FWO project G053221N, G0G3121N, G053221N, G080022N and C14/22/104; and STFC grants ST/P04423/1, ST/V001116/1 and ST/L002868/1; and support from The Swedish Research Council grant 2020-03505.}

\end{document}